\documentclass[conference]{IEEEtran}
\IEEEoverridecommandlockouts
\usepackage{cite}
\usepackage{amsmath,amssymb,amsfonts}
\usepackage{algorithmic}
\usepackage{graphicx}
\usepackage{textcomp}
\usepackage{xcolor}
\usepackage{hyperref}
\usepackage{url}

\usepackage[belowskip=-15pt,aboveskip=4pt]{caption}
\providecommand{\keywords}[1]
{
  \small	
  \textbf{\textit{Keywords---}} #1
}

\def\BibTeX{{\rm B\kern-.05em{\sc i\kern-.025em b}\kern-.08em
    T\kern-.1667em\lower.7ex\hbox{E}\kern-.125emX}}
\begin{document}

\title{Exploring Child-Robot Interaction in Individual and Group settings in India\\
}

\author{Gayathri Manikutty$^{1}$, Sai Ankith Potapragada$^{2}$, Devasena Pasupuleti$^{3}$, Mahesh S. Unnithan$^{4}$, Arjun Venugopal$^{5}$, \\ Pranav Prabha$^{6}$, Arunav H.$^{7}$, Vyshnavi Anil Kumar$^{8}$, Rthuraj P. R.$^{9}$, Rao R Bhavani$^{10}$

\thanks{$^{1}$Gayathri Manikutty, $^{4}$Mahesh S. Unnithan, $^{5}$Arjun Venugopal, $^{6}$Pranav Prabha, and $^{10}$Rao R Bhavani are with AMMACHI Labs, Amrita Vishwa Vidyapeetham, Kollam, Kerala, 690525, India}
\thanks{{\tt\small $^{1}$gayathri.manikutty@ammachilabs.org}}
\thanks{{\tt\small $^{4}$maheshs.unnithan@ammachilabs.org}}
\thanks{{\tt\small $^{5}$arjun.venugopal@ammachilabs.org}}
\thanks{{\tt\small $^{6}$pranav.prabha@ammachilabs.org}}
\thanks{{\tt\small $^{10}$bhavani@ammachilabs.org}}

\thanks{$^{2}$Sai Ankith Potapragada is a student of Department of Computer Science and Applications, Amrita School of
Computing, Amrita Vishwa Vidyapeetham, Amritapuri, India  and doing his internship in AMMACHI Labs, Amrita Vishwa Vidyapeetham, Kollam, Kerala, 690525, India}
\thanks{{\tt\small $^{2}$am.en.p2mca22027@am.students.amrita.edu}}

\thanks{$^{3}$Devasena Pasupuleti is with Department of Systems Innovation, Graduate School of Engineering Science, Osaka University, Japan}
\thanks{{\tt\small $^{3}$deavasena.pasupuleti@irl.sys.es.osaka-u.ac.jp}}

\thanks{$^{7}$Arunav H., $^{8}$Vyshnavi Anil Kumar, and $^{9}$Rthuraj P. R. are students of  Department of Mechanical Engineering, Amrita School of Engineering, Amrita Vishwa Vidyapeetham, Kollam, Kerala, 690525, India and doing their internship in AMMACHI Labs, Amrita Vishwa Vidyapeetham, Kollam, Kerala, 690525, India}
\thanks{{\tt\small $^{7}$am.en.u4are22011@am.students.amrita.edu}}
\thanks{{\tt\small $^{8}$am.en.u4are22036@am.students.amrita.edu}}
\thanks{{\tt\small $^{9}$am.en.u4are22032@am.students.amrita.edu}}
}

\maketitle

\begin{abstract}
This study evaluates the effectiveness of child-robot interactions with the HaKsh-E social robot in India, examining both individual and group interaction settings. The research centers on game-based interactions designed to teach hand hygiene to children aged 7-11. Utilizing video analysis, rubric assessments, and post-study questionnaires, the study gathered data from 36 participants. Findings indicate that children in both settings developed positive perceptions of the robot in terms of the robot's trustworthiness, closeness, and social support. The significant difference in the interaction level scores presented in the study suggests that group settings foster higher levels of interaction, potentially due to peer influence and collaborative dynamics. While both settings showed significant improvements in learning outcomes, the individual setting had more pronounced learning gains. This suggests that personal interactions with the robot might lead to deeper or more effective learning experiences. Consequently, this study concludes that individual interaction settings are more conducive for focused learning gains, while group settings enhance interaction and engagement.
\end{abstract}
\vspace{1em}

\section{Introduction}

Child Robot Interaction (CRI) is a growing field of study that delves into the relationship between children and robots, particularly for therapeutic and educational purposes \cite{coninx, blancas}. We focus our CRI research in an educational setting in India with primary school children. Prior studies have demonstrated that when robots took on roles such as tutors or peers, they helped achieve learning goals \cite{ramachandran, kanda_1} and encouraged a sense of ongoing challenge \cite{wainer} and motivation among the young participants \cite{janssen}. The tutoring experience provided by robots was comparable to and often surpassed, the effectiveness of computer-based tutoring devices, likely because of the embodied nature of the interaction \cite{Leyzberg}. Wijnen et. al noted that children verbalized more and responded more quickly (displayed shorter response time) when interacting with a social robot compared to using a tablet \cite{wijnen}. Recent research has also highlighted that children, typically with ages ranging from 7 to 12 years, not only exhibit positive responses during interactions with robots \cite{nalin} but also demonstrate a high level of trust in them, often surpassing their trust in human interactions \cite{dio}. Literature on CRI studies in educational settings in the global south is scarce to draw conclusions  on broader applicability of robotic technologies in diverse educational systems. A recent study by Ashwini et al. showed successful task completion and positive acceptance of robot-assisted interventions by Indian children aged 3-6 years, indicating the potential effectiveness of such technologies in resource limited educational settings \cite{ashwini}. 

While designing physical space for interactions between children and social robots, the interaction setting could be designed both as a one-on-one setup or a group setup. Prior CRI studies have predominantly focused on examining individual interaction setup between children and robots \cite{straten_1}. Even fewer studies have investigated differences in individual versus group interactions between robots and children. A study by Leite et al. revealed that children demonstrated better recall and semantic details of stories when interacting with robots on a one-on-one basis as opposed to group interaction scenarios though there was no significant difference in learning gains in terms of emotional content of the story \cite{leite}. However, further research is warranted to draw definitive conclusions regarding the effects of interaction settings on learning outcomes and relationship formation between children and robots.

Our research with social robots has been in the domain of behaviour change to promote positive hand hygiene habits among young children with two persuasive robots named Pepe and HaKsh-E in community settings \cite{reference15, reference17, reference18, reference19, reference20, reference21}. HaKsh-E, the social robot we are using in this study, is capable of leading a game-based interaction to teach WHO-recommended hand washing techniques \cite{WHO} to young children. The robot and the child together play a game titled "Land of Hands" wherein their task is to save the princess taken captive by germs due to improper hand washing \cite{reference19}. The robot and the child cooperatively solve three handwashing-related puzzles during gameplay to rescue the princess. In the process of gameplay, the robot helps the child learn about the importance of hand hygiene and good hand hygiene techniques. In an online study that we conducted during the pandemic, we noted that the children showed improved knowledge and ability to demonstrate the WHO-recommended handwashing steps after participating in the game. Also, the pro-social traits of HaKsh-E significantly enhanced interaction and engagement although it did not directly impact the learning outcomes on proper hand hygiene practises.

For educational contexts such as ours, when social robots are deployed in schools, Kanda et al. noted that children tend to interact with the robot in groups \cite{kanda_1, kanda_2}. We have also observed a similar group pattern behaviour among children when we took both Pepe and HaKsh-E robots for pilot interventions in schools. Therefore, for the effective deployment of our robot in schools and rural community centres across India on a long-term basis, we are interested in investigating differences in individual and group interaction with HaKsh-E. To study the group setting, we used a triad interaction with the robot as suggested by Moreland \cite{moreland}. In this study, we will address the following research questions:
\begin{itemize}
    \item What are the differences in engagement in terms of frequency, duration and interaction levels that the children have with HaKsh-E robot in a one-to-one interaction setting versus a group interaction setting?
    \item Does the interaction setting impact children’s learning gain about hand hygiene?
    \item What is the effect of the interaction setting on children’s relationship formation with the robot in terms of perceived social support, trust, and closeness?
\end{itemize}

\section{Methods}
\subsection{Study Tools}
We designed this study as a quantitative CRI study with 36 children. We built on our prior work \cite{reference19} where we evaluated the conversational abilities of the HaKsh-E robot, employing the KindSAR interaction measurement index developed by Fridin et al. \cite{fridin, fridin_2}. This index measures Interaction levels (IL) as a summation of eye contact (indicative of engagement through mutual gaze) and the children's emotional expressions, encompassing facial, body, and vocal reactions during their interactions with HaKsh-E. We used the video data captured during children's interaction sessions and did a post-hoc analysis of the data using the ELAN computer software. ELAN is a professional tool to manually and semi-automatically annotate and transcribe video recordings \cite{elan}. 

We also used the trust, closeness, and perceived social support scales developed by Straten et al. \cite{straten_2} to measure child-robot relationship formation. We added attention checks \cite{abbey} to ensure the reliability of children's responses. Finally, to evaluate children's awareness and learning outcomes related to hand hygiene practices, we created a self-designed questionnaire on hand hygiene awareness.

\subsection{Participants}
36 children participated in this study comprising of 18 girls and 18 boys with ages between 7 and 11 years. All participants were students at Vivekananda Higher Secondary School in Changankulangara, Kerala, India. Convenience sampling was used to select the participants for the study. 

\subsection{Study Setting}
The study started with the registration of participants. Registration included providing voluntary consent by parents for their child to participate in the study and permission to video record the entire session of the robot interacting with the child. After registration, we arranged facilitated sessions in the school with interested participants. We divided the children randomly into two groups. In the first group, 18 children (10 girls and 8 boys) interacted with the robot in an individual one-on-one setting. The second group, consisting of 18 children (8 girls and 10 boys), were divided into six triads. Each triad interacted with the robot in a one-to-many interaction setting. A school teacher escorted the children to the study location one at a time if they were participating in the study individually or in groups of three if they were participating as a triad. 

\begin{figure*}
    \centering
    \includegraphics[width=1\linewidth]{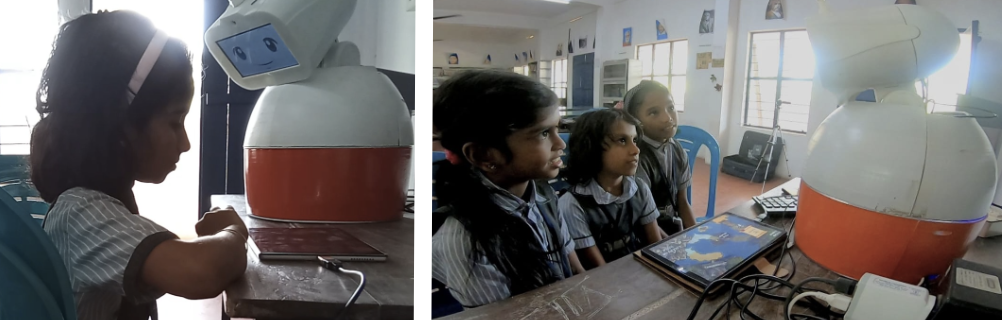}
    \caption{Individual and Group Interaction Sessions}
    \label{fig:indtriadlive}
\end{figure*}

Our study setup is shown in figure where the children sat facing the robot \ref{fig:indtriadlive}. They played the handwashing game on a tablet kept on a table between them and the robot. To ensure the authenticity of the interactions, a researcher discreetly controlled the robot using a Wizard-of-Oz setup throughout the gameplay. There were no inputs given to the children by any human during the gameplay, allowing children to maintain focus entirely on the game and their interactions with the robot. At the start of the study, we administered the learning assessment questionnaire on hand hygiene as a pre-test to each child individually to collect their baseline knowledge on hand washing practice. Immediately after the gameplay, we asked children from both groups to individually rate their perceived trust, closeness, and social support with the robot. We also administered the same learning assessment questionnaire on hand hygiene as a post-test. The learning assessment was completed by each child individually regardless of their interaction setting.

\section{Results and Discussion}
At the outset of the study, we queried participants about their video game-playing habits, previous exposure to robots, and prior interactions with robotic systems. Of the 36 participants, a significant majority (80.5\%, n=29) reported engaging in video game play, with 11.1\% playing games occasionally and a minor 8.3\% (n=3) reporting no engagement with video games at all. The children expressed preferences for a variety of video game genres, although a small subset did not specify a particular preference.

In terms of exposure to robots, the majority of the participants (58.3\%, n=21) revealed that they had never previously encountered a robot. Moreover, a substantial 75\% of the participants indicated that they had never engaged in any form of interaction with a robot. Three students recalled having casual conversations with HaKsh-E earlier during a science and technology outreach activity in their school. Additionally, one student recounted an experience of interacting with a robot as part of a friend's project, and another noted having observed robots speaking in movies.

To remind the readers, the primary objective of this research is to investigate the impact of different participant settings, specifically individual versus group, on interaction levels, learning outcomes and relationship formation between the child and HaKsh-E within an offline game environment. The findings are categorized into three distinct sections: 
The first section is - \emph{Frequency, Duration and Interaction Levels}, which presents the findings about two key factors: the frequency and duration of positive emotional expressions and verbal responses, and the Interaction Level (IL) scores.
The second section is - \emph{Learning Gain}, which presents results from the comparative analysis of the self-designed questionnaire on hand hygiene knowledge obtained from individual and triad participants, before and after the study.
The third section is - \emph{Relationship Formation with the Robot}, which highlights the results of the closeness, trust and perceived social support scales.

\subsection{Frequency, Duration and Interaction levels}
Our first research question was - \emph{what are the differences in engagement in terms of frequency, duration and interaction levels that the children have with HaKsh-E robot in a one-to-one interaction setting versus a group interaction setting?}. 

\subsubsection{Frequency and Duration of Verbal Responses and Positive Emotional Expressions}
We transcribed and categorized the raw utterances and positive emotional expressions of children from the study videos into two categories namely - \emph{verbal responses} and \emph{positive emotional expressions} using the ELAN annotating software \cite{elan}. For verbal responses, we considered children's positive or neutral responses towards the robot or the game. These include "okay", "thank you", etc. We did not consider negative utterances, or utterances where the child appeared confused by saying "huh" or "what?", to serve as a type of social engagement with the robot. Each verbal response was calculated within a five-second time frame, meaning if a child uttered a sentence for 10 seconds, it was considered as two responses, whereas if a child uttered only a word for two seconds, it was considered a single response. 
For positive emotional expressions, we considered only positive expressions such as a child smiling or being surprised towards either the robot or the game, as these expressions have proven to be forms of positive social engagement with robots \cite{ahmad, serholt}. Also, just like we did in our previous study \cite{reference19}, we did not consider any negative emotions such as anger, frustration, etc. to be forms of social engagement although none of the children displayed any negative emotions towards the robot during the study. Similar to the verbal responses, positive emotional expressions were also calculated within a five-second time frame. 

\begin{figure}[htbp]
\centerline{\includegraphics[scale=0.63]{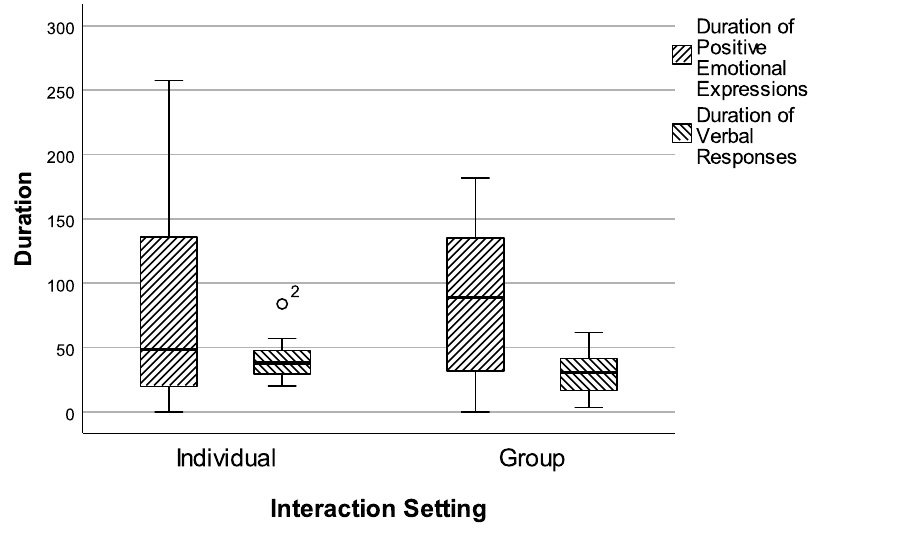}}
\caption{Duration Boxplot}
\label{fig: duration_boxplot}
\end{figure}

\begin{figure}[htbp]
\centerline{\includegraphics[scale=0.7]{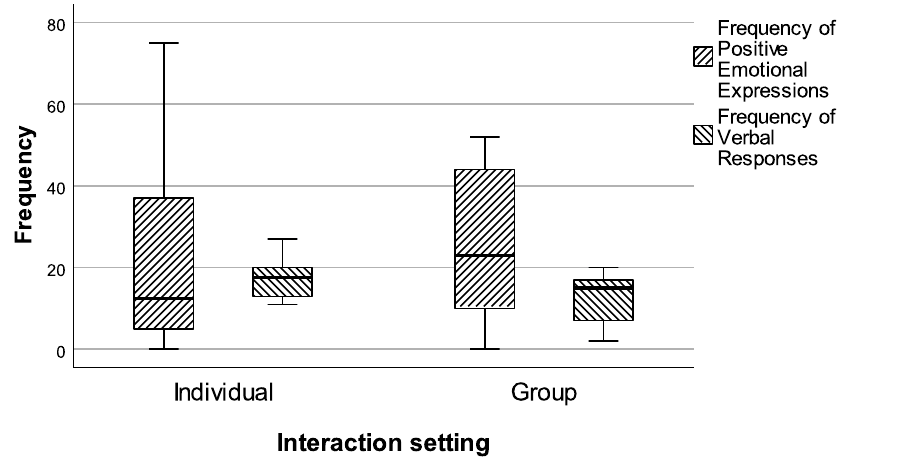}}
\caption{Frequency Boxplot}
\label{fig: frequency_boxplot}
\end{figure}

We have presented frequency, duration and IL as box plots in figure \ref{fig: duration_boxplot} and figure \ref{fig: frequency_boxplot}. Both the frequency and duration of verbal responses and positive emotional expressions were non-normally distributed. Hence, we conducted a non-parametric Mann–Whitney U test ($\alpha$ = 0.05) for all the aforementioned categories in both the individual and group interaction conditions. The results are summarized below.

\begin{enumerate}
    \item \text{Frequency of positive emotional expressions}: While the frequency of positive emotional expressions in the group interaction condition (Median=23, N=18) was higher than that of the independent interaction condition (Median=12.5, N=18), the Mann–Whitney U test showed that this distinction was not statistically significant, U=125.5, z=-1.156, p=0.252.
    
    \item \text{Duration of positive emotional expressions}:
    Again, the duration of positive emotional expressions in the group interaction condition (Median=88.74, N=18) was higher than that of the individual interaction condition (Median=48.37, N=18), but the Mann–Whitney U test indicated that this difference was not statistically significant, U=142, z=-0.633, p=0.542.
    
    \item \text{Frequency of verbal responses}:
    The frequency of verbal responses in the independent interaction condition (Median=17.5, N=18) was higher than that of the group interaction condition (Median=15, N=18) and the Mann–Whitney U test revealed that this disparity was statistically significant, U=94, z=-2.158, p=0.031.
    
    \item \text{Duration of Verbal Responses}:
    The duration of verbal responses in the individual interaction condition (Median=38.05, N=18) was higher than that of the group interaction condition (Median=30.38, N=18) but the Mann–Whitney U test indicated that this difference was not statistically significant, U=106, Z=-1.772, p=0.079.
\end{enumerate}

\subsubsection{Interaction Level (IL) Scores}
As stated earlier, we incorporated the Interaction Level (IL) index developed by Fridin \cite{fridin}. The evaluation of the interaction index at a specific stage "S" within a session is determined by Equation.\ref{eq1}.

\begin{equation}\label{eq1}
    IL_s = EC_s * Sign_s * \sum_{F=1}^{3} W_FF
\end{equation}

In this context, $IL_s$ represents the interaction level, $EC_s$ denotes the extent of eye contact with the proposed platform, and $Sign_s$ is a variable indicating positive or negative interaction. Specifically, $Sign_s$ equals +1 when the child's interaction with the platform is positive, and -1 when the interaction is negative. Furthermore, $F$ represents the Affective Factor and $W_F$ is a binary variable that is set to 1 when the child exhibits an emotional response and 0 when the child does not display any emotional response. Again, none of the children displayed any negative emotional response to the robot.

For our study, the variable EC can take on different values: EC = 3 shows that the child directs their gaze towards the platform (either the robot or the game) during interaction; EC = 1 indicates that the child looks towards the researcher seeking assistance or clarification, and EC = 0 indicates that the child doesn't engage with the platform at all. Also, for the two affective factors, namely facial expressions (especially positive emotional expressions) and verbal responses, the variable F is given a value of 1 if the child displays facial emotions (like smiles and expressions of surprise) and a value of 3 if the child expresses emotions through verbal responses.

A total of 1466 interactions were recorded from the study population (n=36). We categorized the IL data based on the participants' conditions (individual vs. group). Analysis using the Shapiro-Wilk test indicated that the IL data did not adhere to a normal distribution for the participants' conditions. Therefore, we opted to conduct a non-parametric Mann–Whitney U test ($\alpha$ = 0.05) on our dataset. The Mann–Whitney U test showed that the IL index difference was statistically significant between the individual interaction setting (Median=3, N=18) and group interaction setting (Median=3, N=18) U=227506, z=-5.3, \(p<0.001\). 

Thus we conclude that while verbal responses were more frequent in the individual participants setting, there was no significant difference in positive emotional facial expression frequency or duration between the two settings. This suggests that while children may feel more inclined to vocalize their interaction with the robot in the individual setting, their emotional engagement remains consistent across settings. The significant difference in Interaction Level (IL) scores, however, indicates that the group interaction setting fosters higher levels of interaction, potentially due to peer influence and collaborative dynamics.

\subsection{Learning Gain}
Our second research question was - \emph{does the interaction setting impact children’s learning gain about hand hygiene?}. To examine the learning gains through the interaction with HaKsh-E, we created a self-designed questionnaire on hand hygiene awareness with its assessment as a rubric grid. This questionnaire comprised of seven questions based on the content the robot taught the children during the game play. The same seven questions were used for both pre-test and post-test. Each question in the questionnaire was evaluated on a 3-point scale:
\begin{itemize}
    \item Grade 2 denoting ``Independently Answered" - If the child described what they learned from the game without any cues from the researcher, 
    \item Grade 1 denoting ``Verbal Cues were needed" - If the child needed verbal prompts from the researcher to articulate their understanding and 
    \item Grade 0 denoting ``Physical Prompts were needed" - If the researcher had to employ physical cues or verbally tell the answer to assist the child in recalling what was learned from the game.
\end{itemize}

To assess learning gain, two researchers independently watched the video recordings of the students' pre and post-test questionnaires and scored each participant. Cohen's kappa ($\kappa$) was run to determine if there was an agreement between the two researchers’ ratings. There was excellent agreement between the two researchers' ratings in both pre and post-tests with pre-test rating: $\kappa$ = .907 (95\% CI, .858 to .956), \(p < .001\), and post-test rating: $\kappa$ = .867 (95\% CI, .8 to .934), \(p < .001\). All differences in scoring were mutually reconciled by the two researchers.

We followed similar measures adopted by Ramachandran et al. \cite{ramachandran} to calculate normalized learning gain from the intervention so that it accounts for differing levels of knowledge of participants at the beginning of the study. First, the total score for each participant on the pre-test and post-test was calculated by adding up the grades received for each question based on the above rubric and dividing the total by 14, the maximum obtainable grade. The resultant score for each participant ranged between zero and one on the pre-test and post-test. Then the normalized learning gain was calculated as in Equation.\ref{eq2}.

\begin{equation}\label{eq2}
    normalized\ learning\ gain = \frac{score\_post - score\_pre}{1-score\_pre} 
\end{equation}

Normalized learning gain is, thus, the ratio of the raw gain achieved to the possible improvement that could be achieved, as determined by the pre-test score \cite{coletta, hake}. We computed normalized gain for each participant in both interaction settings. In the group interaction setting, one student had a pre-test score of 0.93 but a post score of 0.86. This was because the student got an answer correctly as a lucky guess on the pre-test but she could not answer the same question correctly in the post test. This made her learning gain as -1. As Coletta et al. recommended in their paper \cite{coletta}, we set the learning gain for this student as 0.

We tested the normality of our pre-test data. The pre-test data for the triad group was not normal as assessed by Shapiro-Wilk's test (Individual p = .507, Triad p = .016). Hence the non-parametric Mann Whitney U test ($\alpha$ = 0.05) was used to assess pretest scores between the two experimental groups which showed that the difference in scores were not statistically significant, individual (Median=0.64, n=18) and group (Median=0.64, n=18), U=104.5, z=-1.85, p=.068. The two groups were gender-balanced and there were no significant gender differences in pre-test scores between boys (0.63 $\pm$ 0.166) and girls (0.59 $\pm$ 0.091); t(34) = .809, p=0.424.

\begin{figure}[htbp]
\centerline{\includegraphics[scale=0.38]{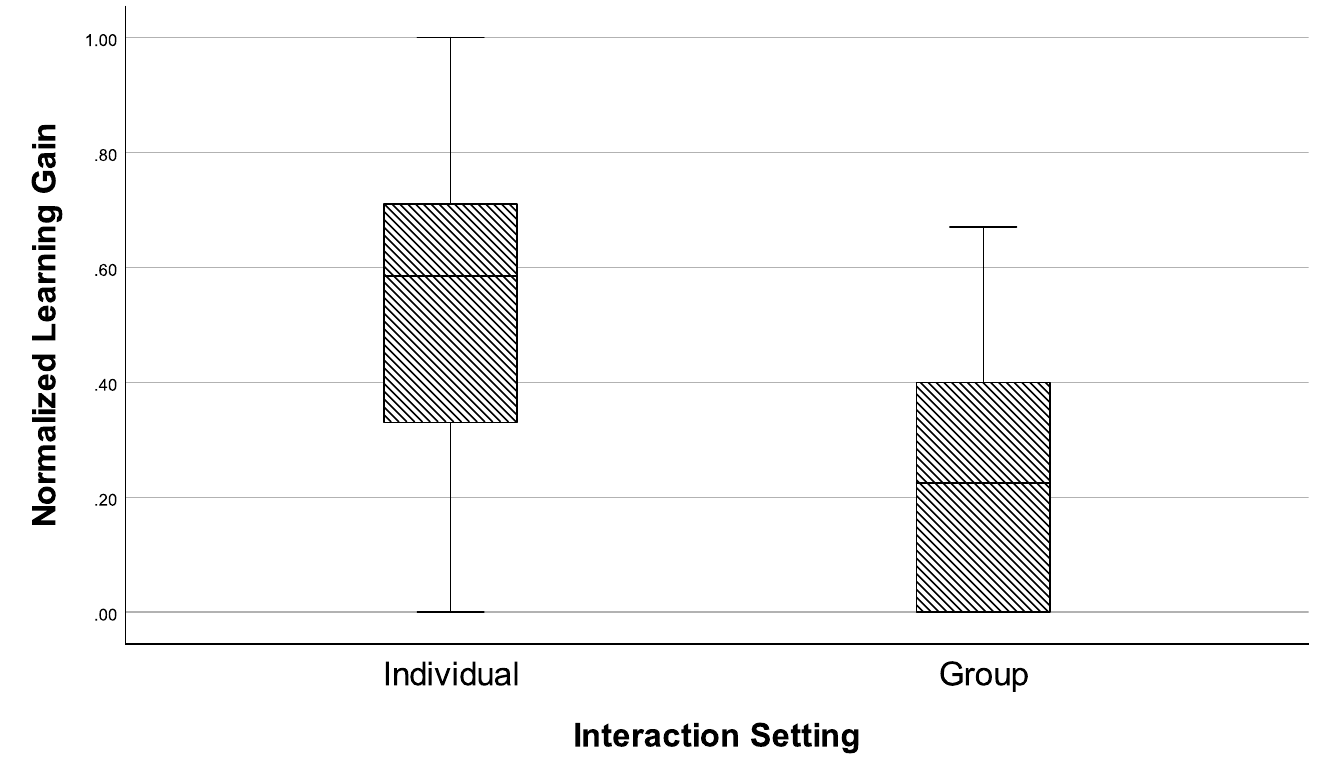}}
\caption{Learning Gain Boxplot}
\label{fig: learning_gain_boxplot}
\end{figure}

\begin{figure}[htbp]
\centerline{\includegraphics[trim={0 0 0 30},clip, scale=0.52]{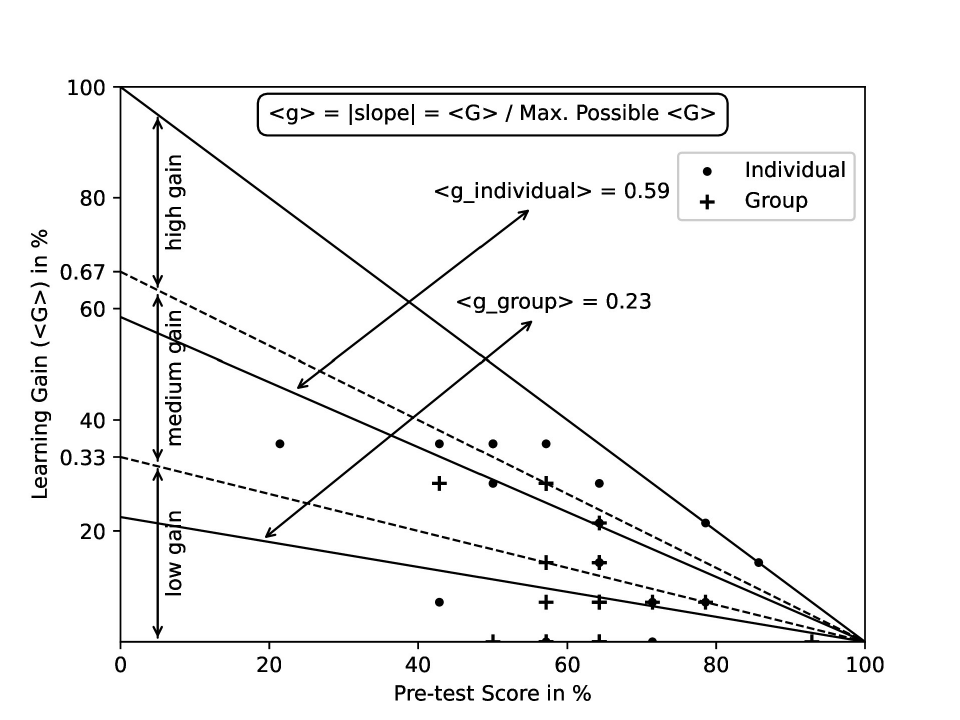}}
\caption{Learning Gain \% vs Pre-test Score \% for Individual and Group}
\label{fig: learning_gain_vs_pre-test}
\end{figure}

When we compared pre-test and post-test data on individual setting, paired samples t-test showed that there was a statistically significant improvement in scores following the intervention with HaKsh-E; t(17)=6.714, \(p<0.001\). Scores improved from 0.598 $\pm$ 0.156 to 0.8 $\pm$ 0.14; an improvement of .205 $\pm$ .13. For the group setting as well, Wilcoxon Signed Ranks test showed a statistically significant improvement in scores (Z=2.99, p=0.003). The median score improved from 0.64 to 0.71 after the intervention. For learning gains, non parametric Mann Whitney U test showed that the difference in learning gains were statistically significant, individual (Median=0.585, n=18) and group (Median=0.225, n=18), U = 81.5, z =-2.565, p = .01. These results indicate that while both interaction settings posted statistically  different between pre and post-test scores, the learning gain for each student was much higher in the individual setting. This agrees with our observation that in the group setting, 6 out of the 18 students posted no change between their pre and post-test scores. The box plot for normalized learning gain is shown in figure \ref{fig: learning_gain_boxplot}. Also, figure \ref{fig: learning_gain_vs_pre-test} shows a plot of learning gain in percentage versus pre-test scores in percentage with group learning gain falling in low gain region and the individual learning gain falling in medium gain region.

To summarize, both individual and group settings resulted in improved learning outcomes, emphasizing the effectiveness of the robot as a tutor for hand hygiene. However, all students do not learn equally well in a group setting; in other words, the individual interaction setting better supports children's learning gains than the group interaction setting.

\subsection{Relationship Formation with the Robot}
\begin{figure}[htbp]
\centerline{\includegraphics[scale=0.39]{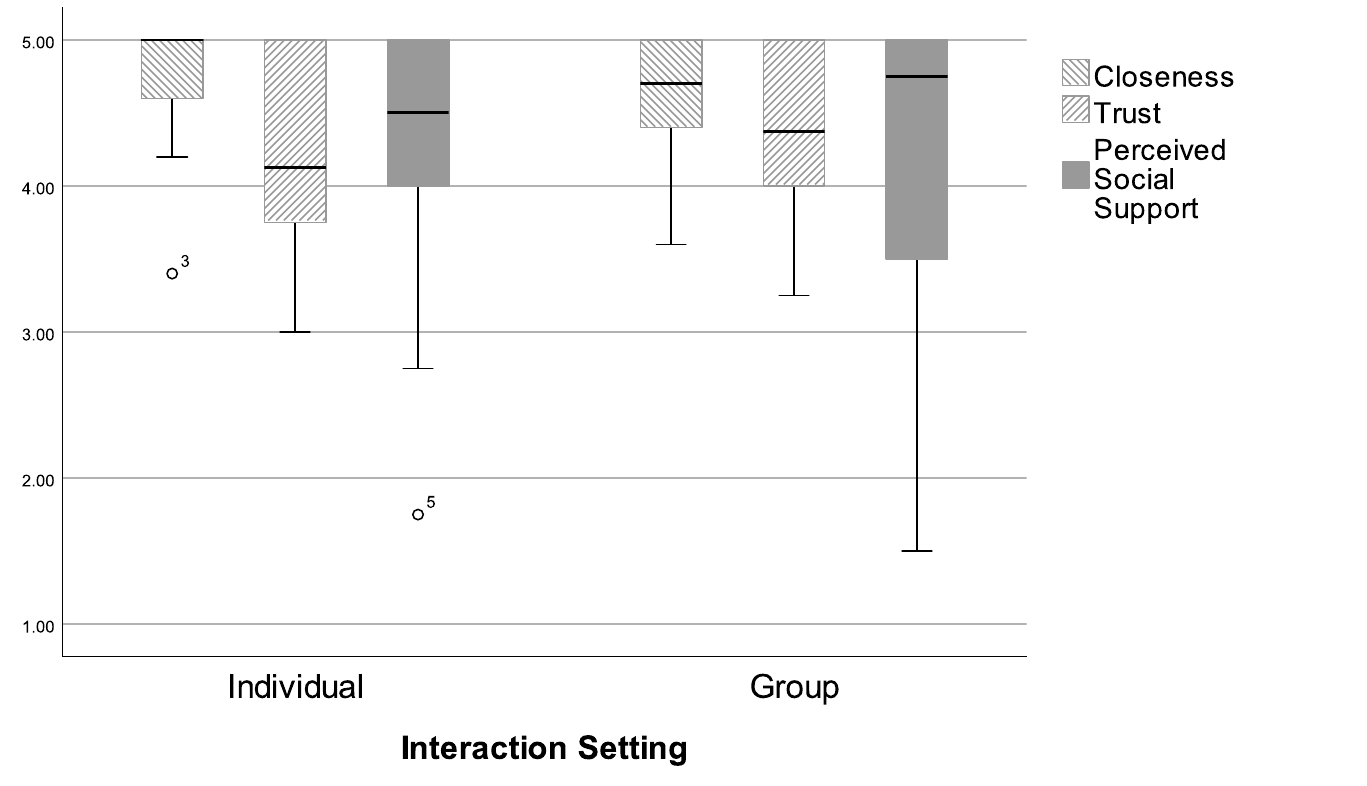}}
\caption{Closeness, Trust, and Perceived Support Boxplot}
\label{fig: closeness_trust_support_boxplot}
\end{figure}

To answer our third research question, which was - \emph{what is the effect of the interaction setting on children’s relationship formation with the robot in terms of perceived social support, trust, and closeness?}, we provided the children with a closeness, trust, and perceived social support questionnaire \cite{straten_2} consisting of 13 items after their gameplay interaction with the robot. The data on three scales was not normal as assessed by the Shapiro-Wilk test except for the trust scale for individual interaction setting (closeness individual \(p<.001\), closeness group p=.003, trust individual p=.072, trust group p=.005, perceived social support individual p=.001, perceived social support group p=.001). The box plot for closeness, trust and perceived social support is shown in figure \ref{fig: closeness_trust_support_boxplot}.

Hence the non-parametric Mann-Whitney U test ($\alpha$ = 0.05) was used to assess differences in the three factors between the two interaction settings. The test showed that the difference in scores was not statistically significant: closeness individual (Median=5.0, n=18) and closeness group (Median=4.7, n=18), U=122, z=-1.374, p=.214. trust individual (Median=4.125, n=18) and trust group (Median=4.375, n=18), U=133, z=-0.945, p=.372. Perceived social support individual (Median=4.5, n=18) and perceived social support group (Median=4.75, n=18), U=159, z=-0.081, p=.938.

Thus we conclude that children's relationship formation with the robot had consistent perceptions of trust, closeness, and perceived social support across individual and group settings. Despite differences in interaction dynamics, children maintained positive perceptions of the robot, indicating a robust sense of rapport regardless of social context. This suggests that social robots hold promise as reliable companions and educators for children, capable of fostering meaningful relationships even in community settings.

Based on the data provided by the children at the beginning of the study regarding prior interactions with robots, only a small fraction of the children had any experience with robots. Most of them encountered robots incidentally in public spaces such as shops or through friends' projects. Consequently, for the majority of participants, the interaction with HaKsh-E represented a novel experience. The novel and intriguing element of interacting with a robot in an educational context provided a unique learning environment that differed significantly from their usual interactions with technology. This new context for learning may have enhanced their receptivity and enthusiasm, contributing to the positive outcomes recorded in this study.

\section{Conclusion}
This research evaluates the effectiveness of HaKsh-E social robot in facilitating game-based interactions with children and explores the interaction levels of child robot interaction and the learning outcomes in one-on-one and group settings. In the individual setting, the frequency of verbal responses was significantly higher, suggesting that children might engage more verbally when they interact one-on-one with the robot. However, the frequency and duration of positive emotional expressions did not differ significantly between settings. The interaction level scores were significantly higher in group settings, indicating that peer influence and collaborative dynamics might enhance overall engagement levels.

We also noted that the interaction setting did impact learning gains. Individual settings resulted in more pronounced learning outcomes, with children showing significant improvements in their understanding of hand hygiene practices. This suggests that personal, focused interactions with the robot facilitate deeper learning, potentially due to the undivided attention and tailored feedback that a one-to-one setting can offer. Although both settings saw improvements, the individual setting was more effective for achieving specific educational goals.

Despite differences in interaction dynamics, the study concluded that children's relationship formation with the robot was not significantly affected by the interaction setting. The perceptions of trust, closeness, and perceived social support towards the robot remained consistently positive across both individual and group settings. This indicates that HaKsh-E was able to establish a stable and positive rapport with children regardless of the setting, suggesting that social robots can be effective communicators and dependable companions in diverse interaction contexts.

Building on the findings of this study on child-robot interaction in educational settings, we plan to conduct longitudinal studies in the future to assess the long-term impacts of robot-assisted learning on children’s hand hygiene habit formation and relationship dynamics with robots. This could provide deeper insights into how sustained interactions influence learning retention, trust, and engagement over time. We are also looking into the effectiveness of HaKsh-E in promoting hand washing behaviour change among rural school children in India as a behaviour change coach. These insights can inform the development of more effective social robots for behaviour change in community contexts in the global south.

\section{Acknowledgement}
The authors express their deep gratitude to Sri Mata Amritanandamayi Devi, a celebrated humanitarian and spiritual leader, for her guidance, support, and inspiration, which significantly contributed to the successful execution of this initiative. Additionally, the authors express gratitude to the teachers and students of Vivekananda Higher Secondary School for their enthusiastic participation throughout the implementation of this research endeavour despite the numerous hiccups encountered during the actual data collection process in the school.


\end{document}